\documentclass[sigconf]{acmart}

\AtBeginDocument{%
  \providecommand\BibTeX{{%
    \normalfont B\kern-0.5em{\scshape i\kern-0.25em b}\kern-0.8em\TeX}}}

\setcopyright{acmcopyright}
\copyrightyear{2018}
\acmYear{2018}
\acmDOI{XXXXXXX.XXXXXXX}

\acmConference[WWW '23]{International World Wide Web Conference}{April 30--May 4, 2023}{Austin, USA} 
\acmPrice{15.00}
\acmISBN{978-1-4503-XXXX-X/18/06}

\usepackage[shortlabels]{enumitem}
\usepackage[utf8]{inputenc}
\usepackage{graphicx}
\usepackage{float}
\usepackage{subfigure}
\usepackage{booktabs}
\usepackage{svg}
\usepackage{color}
\usepackage{makecell}
\usepackage{algorithm}
\usepackage{algorithmic}

\begin{document}
\title{Bubble or Not: Measurements, Analyses, and Findings on the Ethereum ERC721 and ERC1155 Non-fungible Token Ecosystem}

\author{Yixiang Tan}
\authornote{Both authors contributed equally to this research.}
\email{tanyx28@mail2.sysu.edu.cn}
\affiliation{%
	\institution{Sun Yat-sen University}
	\country{China}
}

\author{Zhiying Wu}
\authornotemark[1]
\email{wuzhy95@mail2.sysu.edu.cn}
\affiliation{%
	\institution{Sun Yat-sen University}
	\country{China}
}

\author{Jieli Liu}
\email{liujli7@mail2.sysu.edu.cn}
\affiliation{%
	\institution{Sun Yat-sen University}
	\country{China}
}

\author{Jiajing Wu}
\email{wujiajing@mail.sysu.edu.cn}
\affiliation{%
	\institution{Sun Yat-sen University}
	\country{China}
}

\author{Zibin Zheng}
\email{zhzibin@mail.sysu.edu.cn}
\affiliation{%
	\institution{Sun Yat-sen University}
	\country{China}
}

\author{Ting Chen}
\email{brokendragon@uestc.edu.cn}
\affiliation{%
	\institution{University of Electronic Science and Technology of China}
	\country{China}
}


\begin{abstract}
The non-fungible token (NFT) is an emergent type of cryptocurrency that has garnered extensive attention since its inception. 
The uniqueness, indivisibility and humanistic value of NFTs are the key characteristics that distinguish them from traditional tokens.
The market capitalization of NFT reached 21.5 billion USD in 2021, almost 200 times of all previous transactions. 
However, the subsequent rapid decline in NFT market fever in the second quarter of 2022 casts doubts on the ostensible boom in the NFT market. To date, there has been no comprehensive and systematic study of the NFT trade market or of the NFT bubble and hype phenomenon. To fill this gap, we conduct an in-depth investigation of the whole Ethereum ERC721 and ERC1155 NFT ecosystem via graph analysis and apply several metrics to measure the characteristics of NFTs.
By collecting data from the whole blockchain, we construct three graphs, namely NFT create graph, NFT transfer graph, and NFT hold graph, to characterize the NFT traders, analyze the characteristics of NFTs, and discover many observations and insights. 
Moreover, we propose new indicators to quantify the activeness and value of NFT and propose an algorithm that combines indicators and graph analyses to find bubble NFTs. 
Real-world cases demonstrate that our indicators and approach can be used to discern bubble NFTs effectively.
\end{abstract}


\maketitle

\section{Introduction}

With a market capitalization of more than 1.2 trillion USD and a price of around 68,000 USD per coin in November, 2021 \cite{Bitcoinprice}, Bitcoin has always astounded people with its steadily increasing value and long-term activity since its inception in 2009 \cite{nakamoto2008bitcoin}.
The clout of Bitcoin ushered researchers and investors into blockchain, which is a new technology underpinning cryptocurrencies \cite{chen2020traveling}. Based on this technology, an assortment of tokens were created (e.g. Ether) and their market value peaked at 3 trillion USD approximately on August, 2021 \cite{Cryptocurrencymarket}. 


During the period when the market value of cryptocurrencies skyrocketed, CryptoKitties, the first blockchain-based game, has garnered widely recognized and financial interest in early December 2017 \cite{jiang2021cryptokitties}.
By hybridizing cats with different genes, each new-born cat is unique and, if there is a gene mutation, extremely rare. 
Players in CryptoKitties can trade these cartoon cats with varying monetary values which is based on each of them being unique \cite{serada2021cryptokitties}. 
It can be said that CryptoKitties is the prototype of NFT in the true sense because it has the following characteristics: uniqueness, indivisibility and non-interchangeability. Unlike traditional tokens like Bitcoin and Ether, which are standard coins that all the tokens are equivalent and indistinguishable \cite{wang2021non}, each NFT is unique and cannot be exchanged. 
Additionally, due to indivisible nature of NFTs, buying 0.1 token, which occurs frequently in Bitcoin trade, is not permitted. 
In contrast to the conventional token such as Bitcoin which is just a name, NFTs exhibit more humanistic values because they contain more information and serve as a culture symbol.

Ethereum, a platform that issues Ether with the second largest market value in cryptocurrency and is currently the largest NFT trading platform, provides two main protocols for creating NFTs: ERC721 and ERC1155. 
As with ERC1155, ERC721 is a token standard that defines an interface to allow NFT to be managed, owned, and traded by a smart contract \cite{bella2022blockchains}. 
However, there are a large quantity of differences in the ways of creating and transferring NFTs between them. 
ERC721 needs to create a new smart contract (e.g., Cryptokitties contract) to create a new kind of NFT, whereas ERC1155 can deploy infinite kinds of NFT in one smart contract. 
Moreover, ERC721 permits only one-NFT transfer in a transaction whereas a batch of NFTs being transferred in a transaction is sanctioned in ERC1155 which can save many gas fees. 
These features of the two protocols make the corresponding trading, the characteristics of NFTs, and trading participants diverse in many ways.

Undoubtedly, uniqueness, non-interchangeability and humanistic values which are the merits of NFT and systematic protocols make the creation and trade of NFT a hit. 
As the market capitalization of traditional tokens has increased steadily in the initial stage of its development, the market capitalization of NFTs has jumped from only 70 million USD to near 25 billion USD in 2021 which is called ``the first year'' of NFT \cite{NFTmarket}.
Meanwhile, the market capitalization of cryptocurrencies reached a plateau at 2.5 to 2.9 trillion USD in November 2021. 
Just when it seemed that cryptocurrencies were very prosperous, the subsequent consecutive plummet made their market capitalization one-third than its peak. The same to the market capitalization of NFT, after it peaked at around 36 trillion USD in January 28th 2022, a fluctuation appearing and brought it to, fortunately, 23 trillion USD in September 5th 2022. 
No one knows whether the fate of NFT is similar to cryptocurrencies, nevertheless, Beeple, who created the most valuable (more than 69 million USD) NFT called ``Everydays : The First 5000 Days'' \cite{Beeple} said NFT art is absolutely a bubble and exchanged all his Ether to USD \cite{Beeplesays}. 
As the clout of NFT increasing, the dark side of NFT was uncovered which included unauthorized NFTs, wash trading and scams. 
For example, a CryptoPunk NFT was sold for 532 million USD in December 2021 and it was proved to be traded between many wallets of a single user \cite{Darkside}. Some rumors mixed with real events were reported by the mindless media, making the NFT even more elusive.

\begin{figure}[t]
    \centering
    \includegraphics[width=8cm]{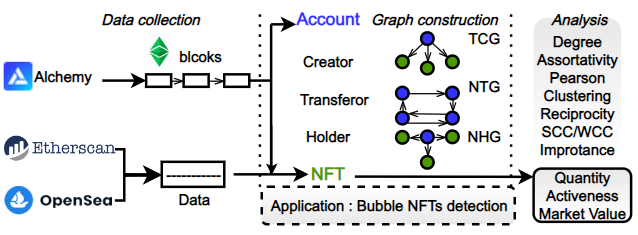}
    \caption{An overview of our framework.}
\end{figure}

Consequently, some questions come up and should be solved urgently: Is NFT truly (or to what extent) a bubble hidden behind the prosperous facade, and how to discern the bubble in the trading ecosystem of NFT? 
Unfortunately, little is known about the ecosystem of NFT because most of the studies \cite{chen2020understanding,chen2020traveling} focus on traditional tokens or just analyze the characteristics of users and contracts, not the NFTs \cite{lee2020measurements}. 
Furthermore, measuring the activeness of NFTs is a challenging task due to the fact that there is only one NFT with the same token ID, which is significantly less than the traditional tokens. 
Assorted humanistic values contained in NFTs also make the trend of their prices much more intangible than traditional tokens, which are just a name. 
In an attempt to fill the gap in research and disclose the characteristics of NFT ecosystem, this paper proposes an approach for analyzing the ERC721 and ERC1155 NFT ecosystems. 
The frame of our work can be seen in Figure 1, we divide our work into four phases.
The data collection phase, which is the first phase, collects all NFT transaction records and event logs on Ethereum. 
By analyzing this data, we classify the actions in NFT trade into three categories chronologically: creation, transfer, and hold. 
Then, we construct three graphs to model the NFT trading, i.e., NFT creation graph (NCG), NFT transfer graph (NTG) and NFT hold graph (NHG).
In the third phase, we conduct a systematic analysis of the three graphs and extract new findings from them. 
Moreover, by analysing the NFT trade data, we have an overview of the behavior of NFTs and propose new indicators to measure them. 
Based on the statistics and indicators, we propose a new approach to detect wash trading issues in the NFT trade network finally.

In summary, we make the following contributions.
\begin{enumerate}[-,topsep=0pt]
    \item[(1)] To the best of our knowledge, we are the \emph{first} to conduct a comprehensive investigation on the whole Ethereum ERC721 and ERC115 NFT ecosystem. We outline the characteristics of NFT traders via graph analysis and propose new NFT indicators to quantify their activeness and value. We also summarize the trends of some quantitative criteria of the whole NFT ecosystem.
    \item[(2)] Using graph analysis and other methodologies, we acquire novel observations and findings about both NFT traders and NFTs in the Ethereum NFT ecosystem. They can help us gain a more comprehensive understanding of this ecosystem. In particular, we find that some anomalies like automatic programs, scam projects, and wash trades also exist. 
    \item[(3)] By combining the new indicators with graph analysis, we propose an algorithm to detect the bubble NFTs. The reported cases show the feasibility and effectiveness of our algorithm. We will release all the relevant data and codes after publication.
\end{enumerate}
The rest of the paper is organized as follows. After reviewing related work in Section 2, we detail the data collection method in Section 3. 
Section 4 answers 3 important questions about the users in the NFT trade network. 
Section 5 analyzes the characteristics of NFTs and proposes some new indicators to measure their activeness. 
After presenting the new approach for detecting bubble NFTs mainly based on the two new indicators in Section 6, we conclude the paper and discuss future work in Section 7.
\section{Related work}
Since the creation of Bitcoin, several works exploring cryptocurrencies and blockchain networks have emerged. 
There are three types of research closely related to our work. 
The first type, which is also the most relevant to our study, is the economic analysis of the cryptocurrency market. 
Paper \cite{chen2020traveling} performed a comprehensive understanding of the whole ERC20 token ecosystem by analyzing more than 160,000 tokens. 
The criteria of a successful token were described by the authors of the papers \cite{conley2017blockchain,howell2020initial}. 
It is worthy mentioning that paper\cite{ante2022non} analyzed the NFT market by examining the number of NFT sales, NFT trade volume and the number of unique blockchain wallets that traded NFTs. 
Nevertheless, the authors collect data from only 14 NFT projects, which is insufficient to enclose the characteristics of the NFT market. 
The second type focuses on the analysis of blockchain networks such as Bitcoin \cite{ermann2018google,ron2013quantitative,spagnuolo2014bitiodine} and Ethereum \cite{ferretti2020ethereum,somin2018network,somin2018social,victor2019measuring}. 
Papers \cite{chen2020understanding,chen2020traveling,lee2020measurements} conducted a systematic analysis of the network they constructed and obtained characteristics of the Ethereum ecosystem. 
Note that these papers just analyzed the characteristics of the users, not the tokens.
Furthermore, the trends in the number of users and transactions are not stated, which could indicate the activeness of the token market. 
The last type mainly proposed new approaches to detect or handle security issues, e.g. revealing Ponzi schemes \cite{bartoletti2018data,bartoletti2020dissecting,chen2018detecting}, the security of smart contract \cite{atzei2017survey,howell2020initial,torres2018osiris}, and wash trade detection \cite{xia2021trade,von2022nft}. 
Different from those papers above, our paper examines the whole Ethereum NFT trading network of more than 74 million tokens in an effort to get a thorough knowledge of the NFT ecosystem. 
\section{Data collection}
Our data includes NFT transfer records, category labels, and NFT descriptive data. 
We launch Alchemy\footnote{https://www.alchemy.com/}, an Ethereum data service, to invocate RPC (Remote Procedure Call) interfaces to extract token transfer records on Ethereum before 1,500,000 blocks (from block \#0 to block \#1,499,999) in total, with a time span from July 2015 to June 2022. 
Specifically, we get the event logs on Ethereum through the RPC interface $eth\_getLogs$, and filter the token transfer about NFT in terms of the ERC721 \cite{eip721} and ERC1155~\cite{eip1155} standards.
Per token transfer record include the address of transferors, recipients and contracts, block-number, timestamp, token ID, transaction hash and values which only exist in ERC1155. 

By crawling data in Etherscan\footnote{https://etherscan.io}, which is the largest blockchain browser, we get the label of NFT categories in the trade network. 
We also collect descriptive data of NFTs in the platform Opensea\footnote{https://opensea.io}, which is the largest NFT marketplace.
The categories of NFTs are art, collectibles, ENS, music, sports, gaming and decentraland. 
The NFT descriptive data are a group of data including name, total supply, total volume, address, created date, description and so on. 
Table \ref{tab:data} shows the specific information of the data we collected.
\begin{table}[t]
  \caption{The description of our NFT Dataset on Ethereum}
  \label{tab:data}
  \begin{center}
    \begin{tabular}{l|r} 
      \toprule
      \textbf{Data type} &  \textbf{Data number}\\
      \hline
      ERC721 transfer records  & 121,762,287\\
      ERC1155 transfer records & 12,313,012\\
      NFT descriptive text items & 1,447\\
      NFT category label & 15,983,402\\
      \bottomrule
    \end{tabular}    
  \end{center}    
\end{table}

In addition, we deploy our dataset on Galaxybase \cite{galaxybase}, a distributed graph database, with 3 cloud computation nodes where each node is equipped with 8 vCPUs, 64GB RAM.

\section{Characteristics of NFT traders}
In order to conduct a specific and precise analysis on NFT Traders, the NFT traders are divided into three categories: creators, transferors, and holders, which represent the participants of NFT trade procedures in chronological order. 
In this section, we address three important questions about the characteristics of NFT traders in order to provide a thorough description of the NFT trade network. 
To answer these questions, in the following parts, we focus our study on graph definition, construction, and analysis.
Based on the analysis results, we arrive at the following findings:
\begin{enumerate}[-,topsep=0pt]
\item[$\bullet$]\textbf{Finding 1.} A small number of accounts created a large number of NFTs, a large number of accounts created a small number of NFTs. People prefer to using ERC721 to create NFTs than ERC1155 though the latter is more effective.
\item[$\bullet$]\textbf{Finding 2.} A small number of accounts are involved in a large number of NFT transfers and a large number of accounts are involved in a small number of NFT transfers indicating that the majority of accounts are not active. \item[$\bullet$]\textbf{Finding 3.} A major proportion of NFTs are possessed by the leading accounts, implying that the leading accounts may be able to dominate the whole NFT market. 
\item[$\bullet$]\textbf{Finding 4.} The number of NFT creators, transferors, and holders is generally shown to be on an upward trend. However, there may be some automatic programs involved.
\end{enumerate}
\subsection{Who Create These NFTs?}
When an NFT is created, it joins the NFT ecosystem, and this is the first phase of the NFT trade, which is followed with interest.
So, we want to know some traits of the NFT creators, including the numbers (of creators and of created NFTs) and the trend of these numbers. 
However, the NFT ecosystem is of an anonymous nature, and it is impossible to expose the identity of an NFT creator owing to the capacity to create an NFT using only an address on Ethereum. Therefore, we consider an address as a creator in order to construct the NFT creation graph. 
For clarity, the terms "address" and "account" are used interchangeably throughout this paper.

\textbf{NCG Definition and Construction.} 
\emph{NCG = (V, E)}, where $V$ is a set of nodes that the outdegree nodes are accounts and the indegree nodes are NFT IDs and the \emph{E} is a set of directed edges with its attribute \emph{t} which is a timestamp, indicating the creating time. 

\begin{figure}[t]
    \subfigure[NCG]{
        \includegraphics[width=0.47\linewidth]{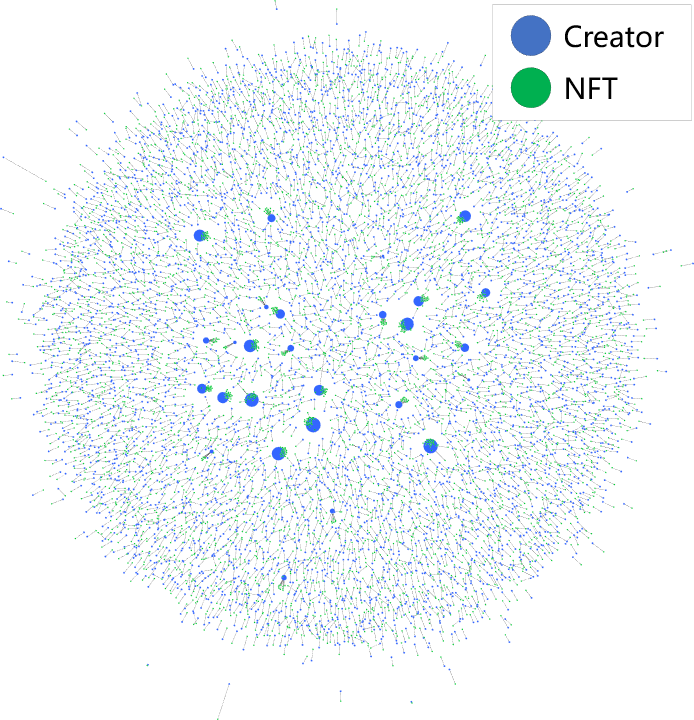}
        \label{fig:ncg}
    }
    \subfigure[Outdegree distribution of NCG]{
        \includegraphics[width=0.47\linewidth]{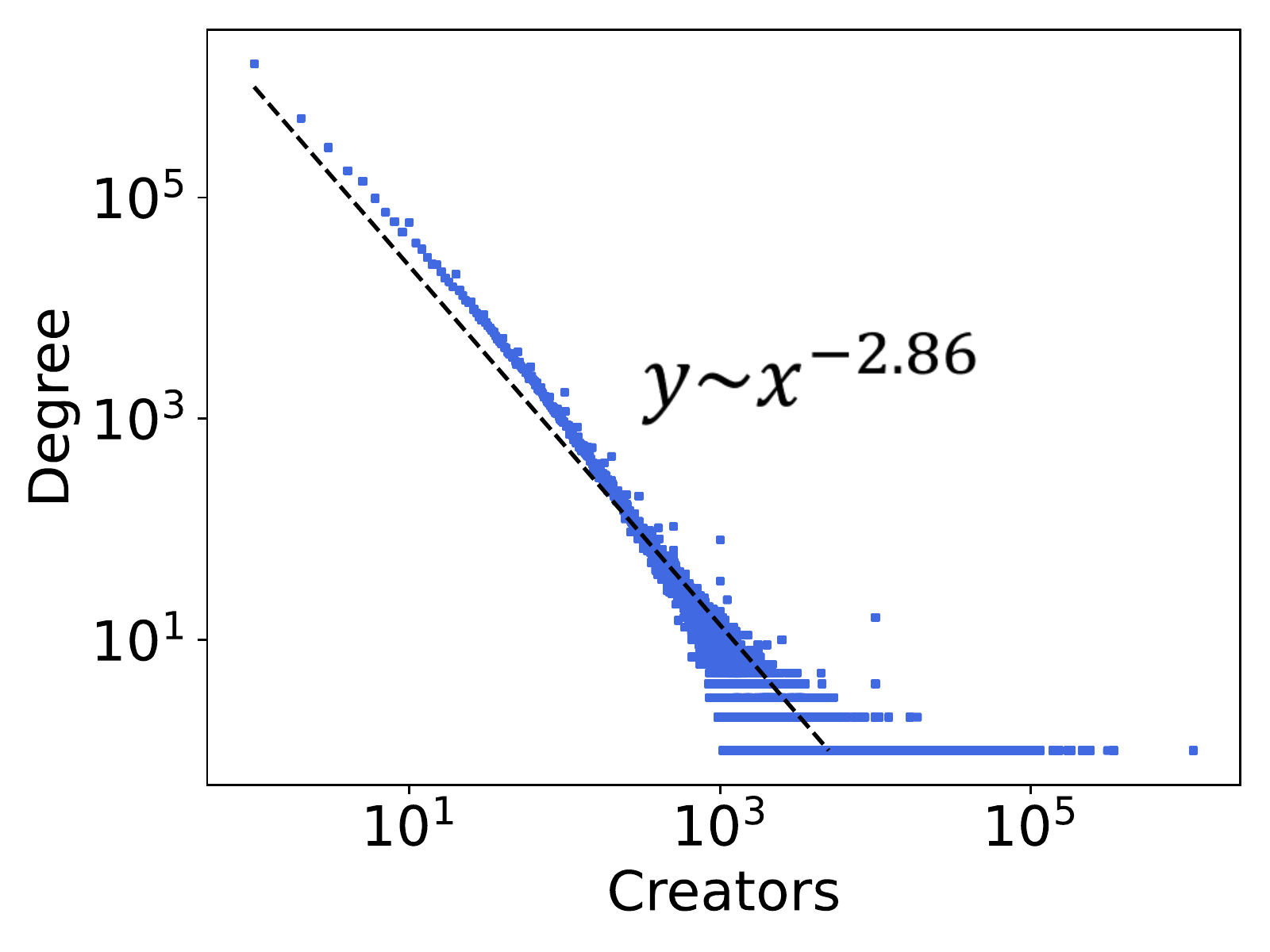}
        \label{fig:ncg_plot_degree_dist}
    }
    \subfigure[The number of NFT creators]{
        \includegraphics[width=0.47\linewidth]{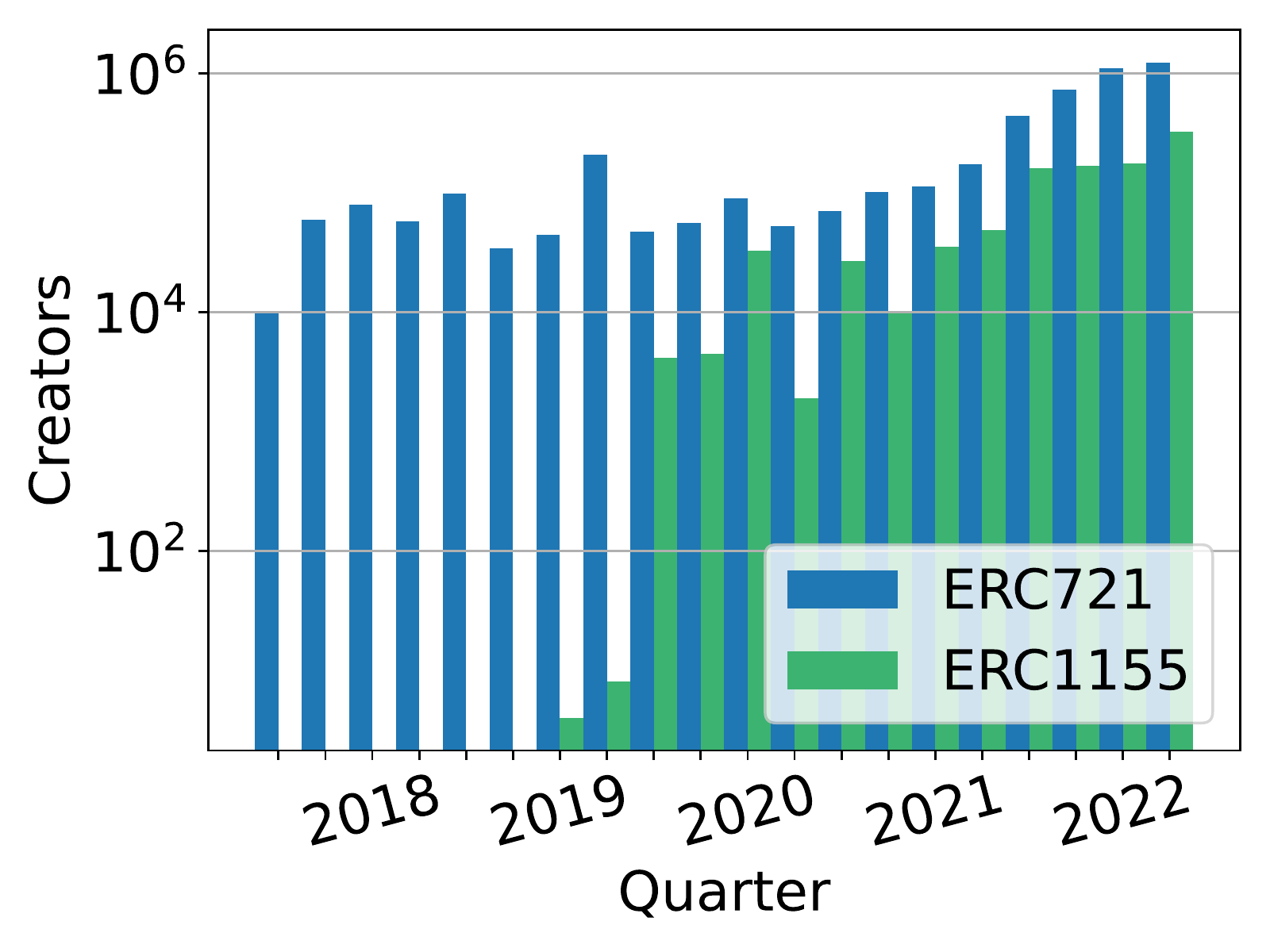}
        \label{fig:ncg_plot_tend_1}
    }\subfigure[The number of created NFT]{
        \includegraphics[width=0.47\linewidth]{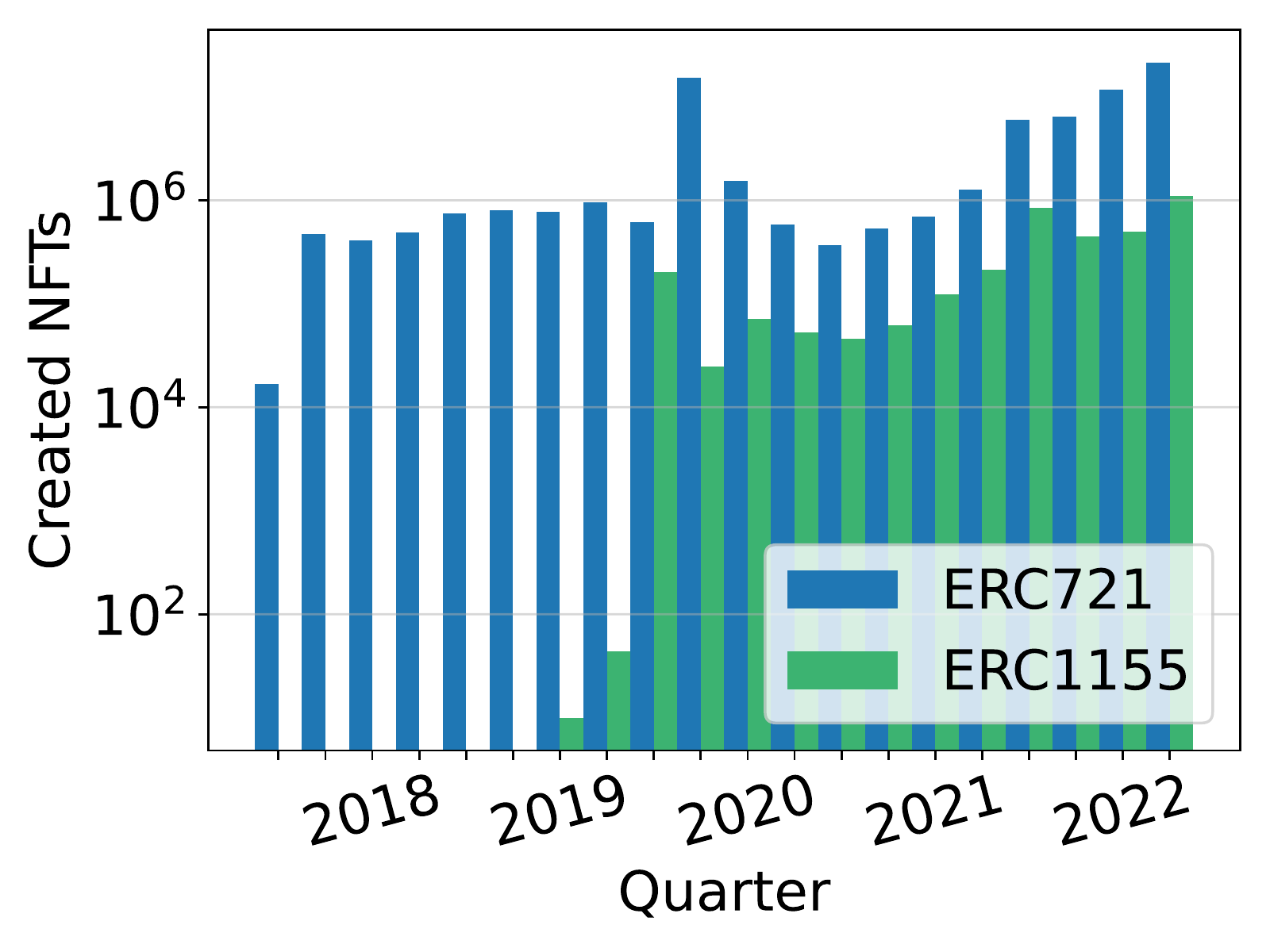}
        \label{fig:ncg_plot_tend_2}
    }
    \caption{Visualization and analysis of NCG.}
\end{figure}

NCG has 78,471,516 nodes and 74,825,956 edges which means around 3.65 million accounts created about 75 million NFTs. 
However, only 3,682,417 ERC1155 NFTs are created comparing to the 71,143,539 ERC721 NFTs. 
To have an overview of the NCG, we randomly choose 10,000 edges with different colors of the two nodes and show the result in Figure \ref{fig:ncg}. 
The blue nodes denote the creators, and the green nodes mean the NFT created by the creator. 
The size of the nodes indicates the number of NFTs that the creators created. 
Note that some accounts created a large quantity of NFTs, which is surprising and anomalous. 
So we plot the outdegree distribution of the NCG in Figure \ref{fig:ncg_plot_degree_dist} for further analysis. 

As can be seen, the outdegree distribution follows a power-law distribution, with a few large-outdegree nodes and numerous small-outdegree nodes, which means a few accounts create the majority of NFTs.
The fitted line $y \sim x^{-\alpha}$ for the distribution is plotted. 
Nearly half (44\%) of the creators only created one NFT.
Only 1\% of creators created 233 NFTs, while 90\% created fewer than 21 NFTs, demonstrating that most creators are inactive.
Surprisingly, the account creating the most number of NFTs is 0x28\footnote{0x283af0b28c62c092c9727f1ee09c02ca627eb7f5}, who created 1,113,140 NFTs accounting for only 1.5\% total NFTs. 
We find 0x28 is the Ethereum Name Service (ENS) whose NFT is a name ending with ".eth". 
It is worthy noting that the address creating the top 20 number of NFTs (about 98,012) is 0x3d\footnote{0xd387a6e4e84a6c86bd90c158c6028a58cc8ac459} which belongs to a famous NFT collector named Pransky with a lot of followers on Twitter.
This information suggests that, while many creators only generate a few NFTs, there are certain people who are dedicated to NFT creation and appeal to others to get involved in it. 
Nevertheless, we find that there are also some bubbles in the creators.
The address 0xe4\footnote{0xe4a8dfca175cdca4ae370f5b7aaff24bd1c9c8ef}, named ``flashbots-builder.eth'', creating massive NFTs automatically as the $5^{th}$ most NFTs creator.
It is no hard to guess that how many NFTs are not created by real users.

Trends in the number of creators and the number of created NFTs are also crucial indicators of the future development of NFTs. 
Due to the different characteristics of the two protocols i.e. ERC721 and ERC1155, we calculate the number of creators who have used these two protocols to create NFTs (Figure \ref{fig:ncg_plot_tend_1}) and the number of created NFTs (Figure \ref{fig:ncg_plot_tend_2}) by quarter separately using the timestamp recorded in the edges of NCG. 
Since the birth of ERC721, the number of creators who used ERC721 to create NFTs skyrocketed from 100 to one million per quarter, then fell to around 100,000, and finally witnessed a steady uptrend to more than one million in the second quarter of 2022. 
Unexpectedly, despite the fact that ERC1155 can produce limitless kinds of NFTs in a single smart contract, the number of creators who use it to make NFTs is substantially lower than that of ERC721 users, suggesting the characteristic "uniqueness" is better embodied in ERC721 NFTs.
Likewise, the trend in the number of new-created NFT kinds soared in the first phase, then dipped somewhat before consistently increasing till today. 
The result infers that NFT will be created more and more and that the potential of ERC1155 has not been developed yet.


\subsection{Who Take Part in the NFT Trading?}
After being created, the NFT may be traded by users, manifesting the activity of the whole NFT ecosystem. 
In this section, we discuss the NFT transferor characteristics by constructing and analyzing the NFT transfer graph (NTG).

\textbf{NTG Definition and Construction.} 
\emph{NTG = (V, E)}, where \emph{V} is a set of nodes representing accounts and \emph{E} is a set of directed edges representing NFT transfers between accounts, with \emph{w} being the weight corresponding to the transfer number of times, indicating the activeness of the participants.

\begin{figure}[t]
    \centering
    \includegraphics[width=6cm]{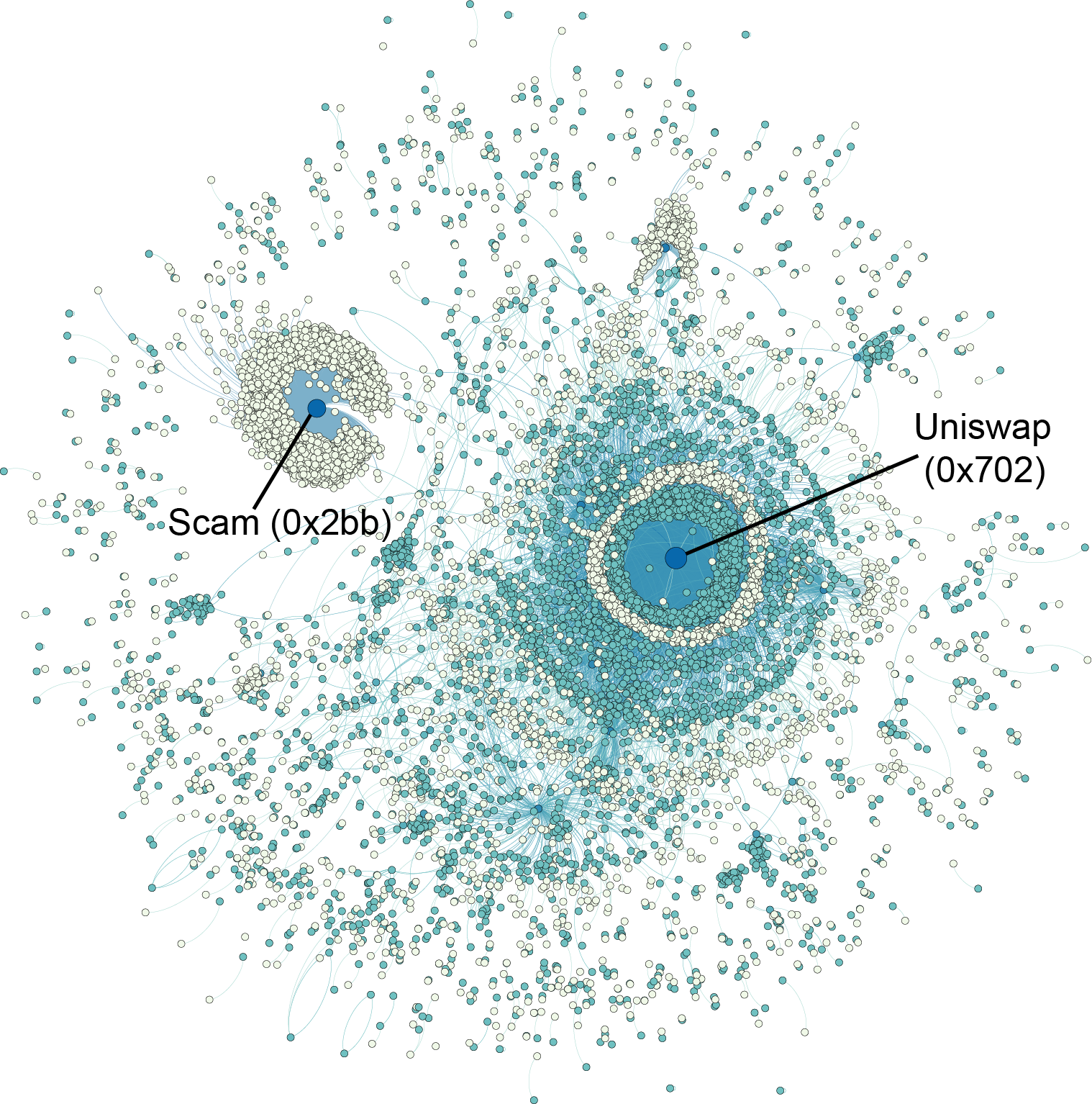}
    \caption{The visualization of NTG.}
    \label{fig:ntg}
\end{figure}

By analyzing the number of nodes and edges of NTG, we know that 5,078,013 accounts transfer NFTs 56,641,999 times, where ERC 721 NFTs are transferred 48,429,314 times by 4,442,408 accounts and ERC1155 NFTs are transferred 8,212,685 times (17\% of ERC721) by 1,433,861 (32\% of ERC721) accounts. 
The average transfer number of ERC721 is 10.9, while the number of ERC1155 is only 5.7. The reason may be that ERC1155 allows to transfer several NFTs in a single transfer.
Figure \ref{fig:ntg} shows a sampled NTG with 10,000 edges, where the size and color depth of the node grow with the number of NFTs the corresponding account transfers.
As can be seen, there are two blue nodes with plenty of connections, which attract our special attention. 
We find that the right bigger one is a Uniswap, and the left one is an NFT scam project with the address 0x2b\footnote{0x2bb5454c0da5f2c2c3cfe81fdedbe630048376c7}.
This account sends advertisements to other accounts by transferring valueless NFTs so its outdegree of is much higher than its indegree. 

\begin{figure}[t]
    \subfigure[Indegree distribution of NTG]{
        \includegraphics[width=0.47\linewidth]{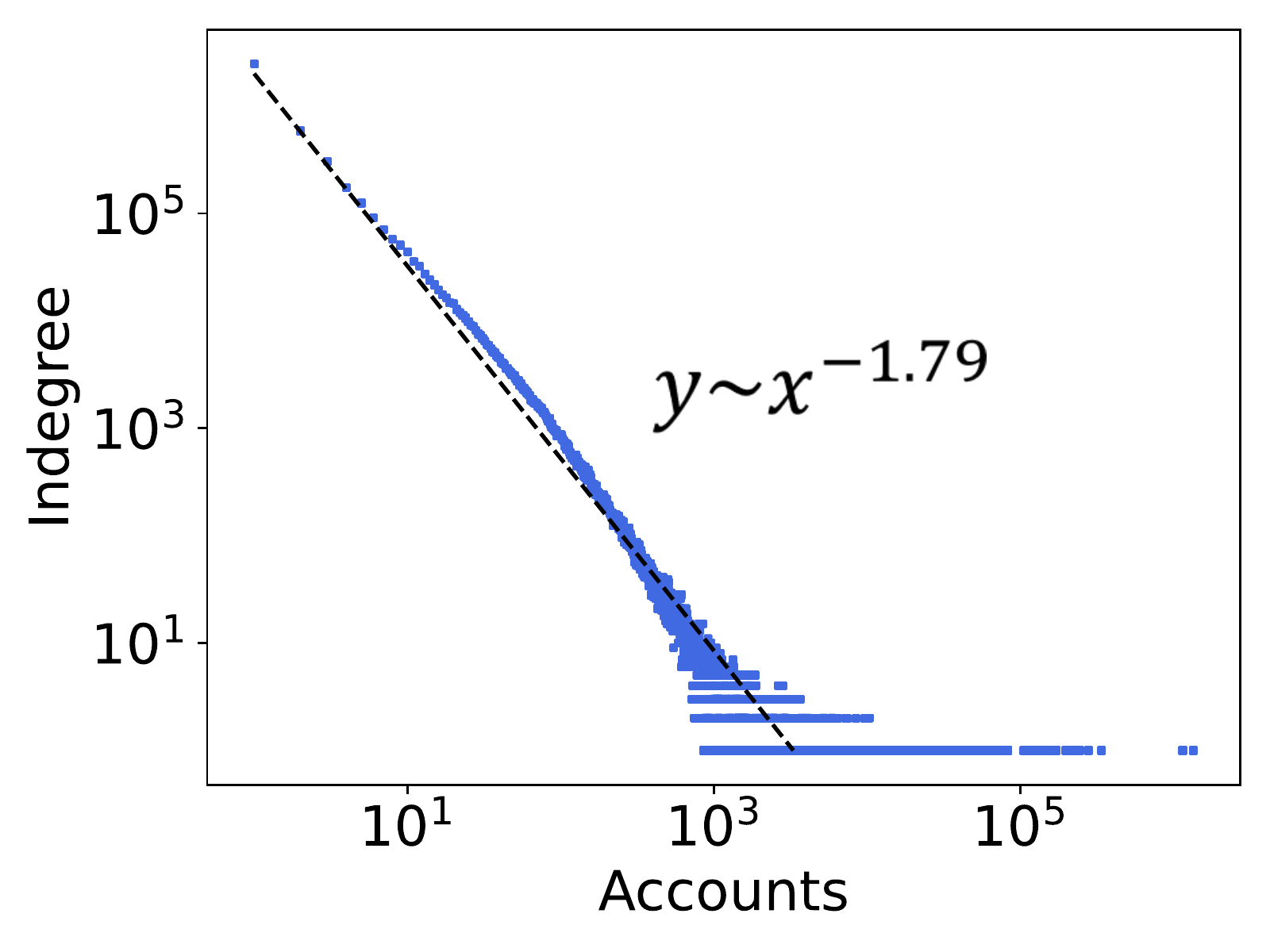}
        \label{fig:ntg_plot_indegree_dist}
    }
    \subfigure[Outdegree distribution of NTG]{
        \includegraphics[width=0.47\linewidth]{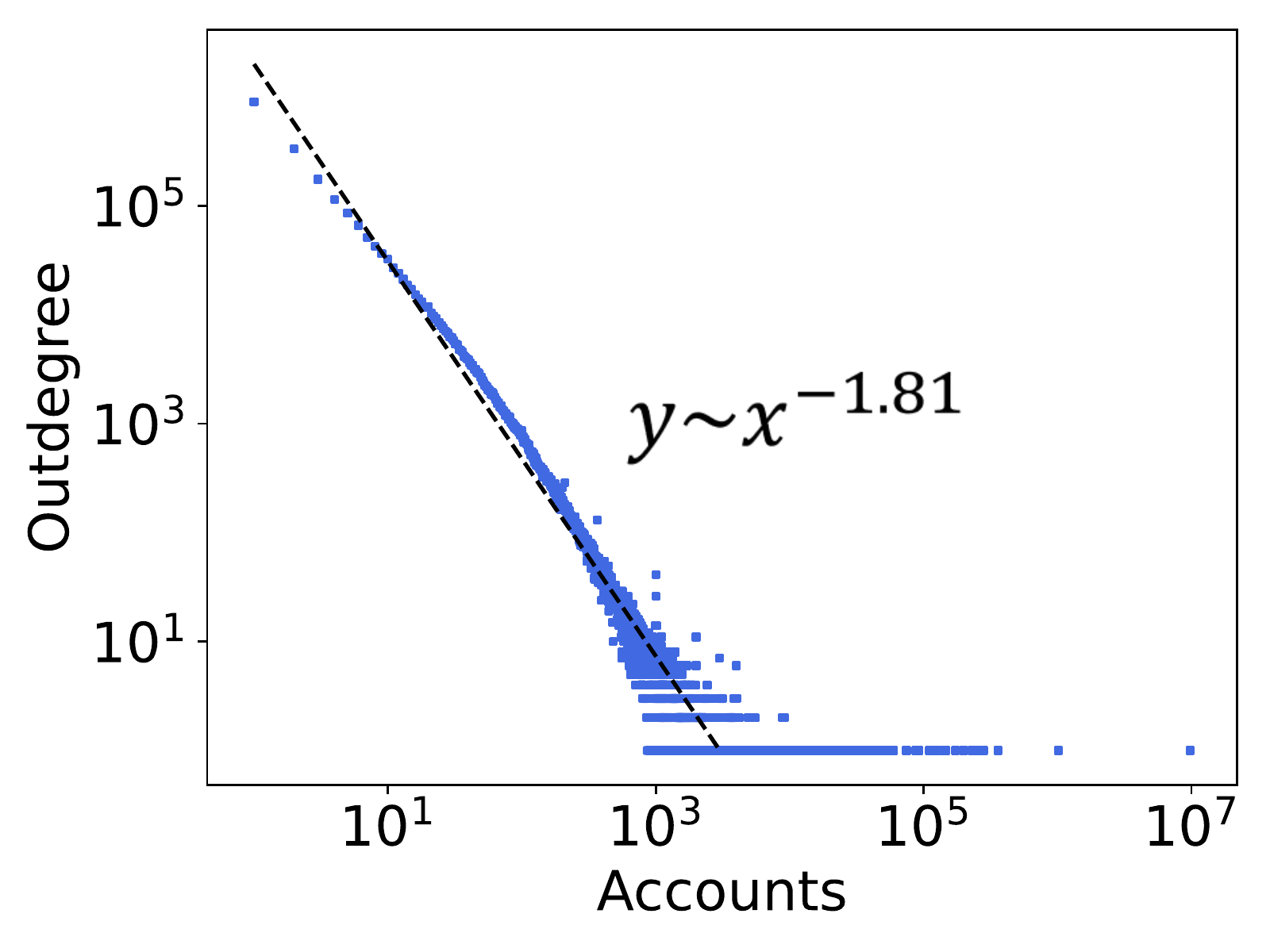}
        \label{fig:ntg_plot_outdegree_dist}
    }
    \subfigure[The number of NFT transferors]{
        \includegraphics[width=0.47\linewidth]{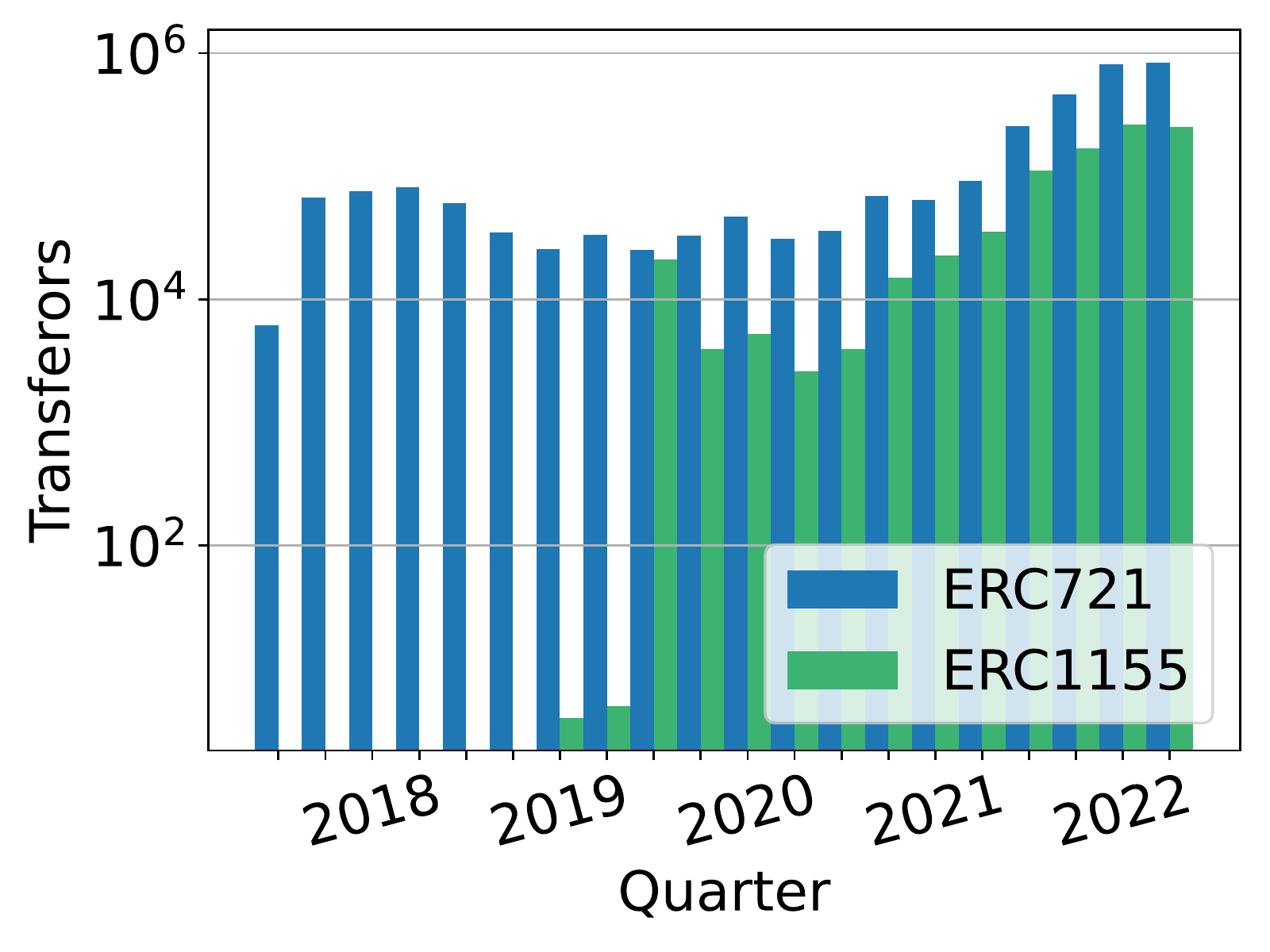}
        \label{fig:ntg_plot_tend_1}
    }
    \subfigure[The number of NFT transfer times]{
        \includegraphics[width=0.47\linewidth]{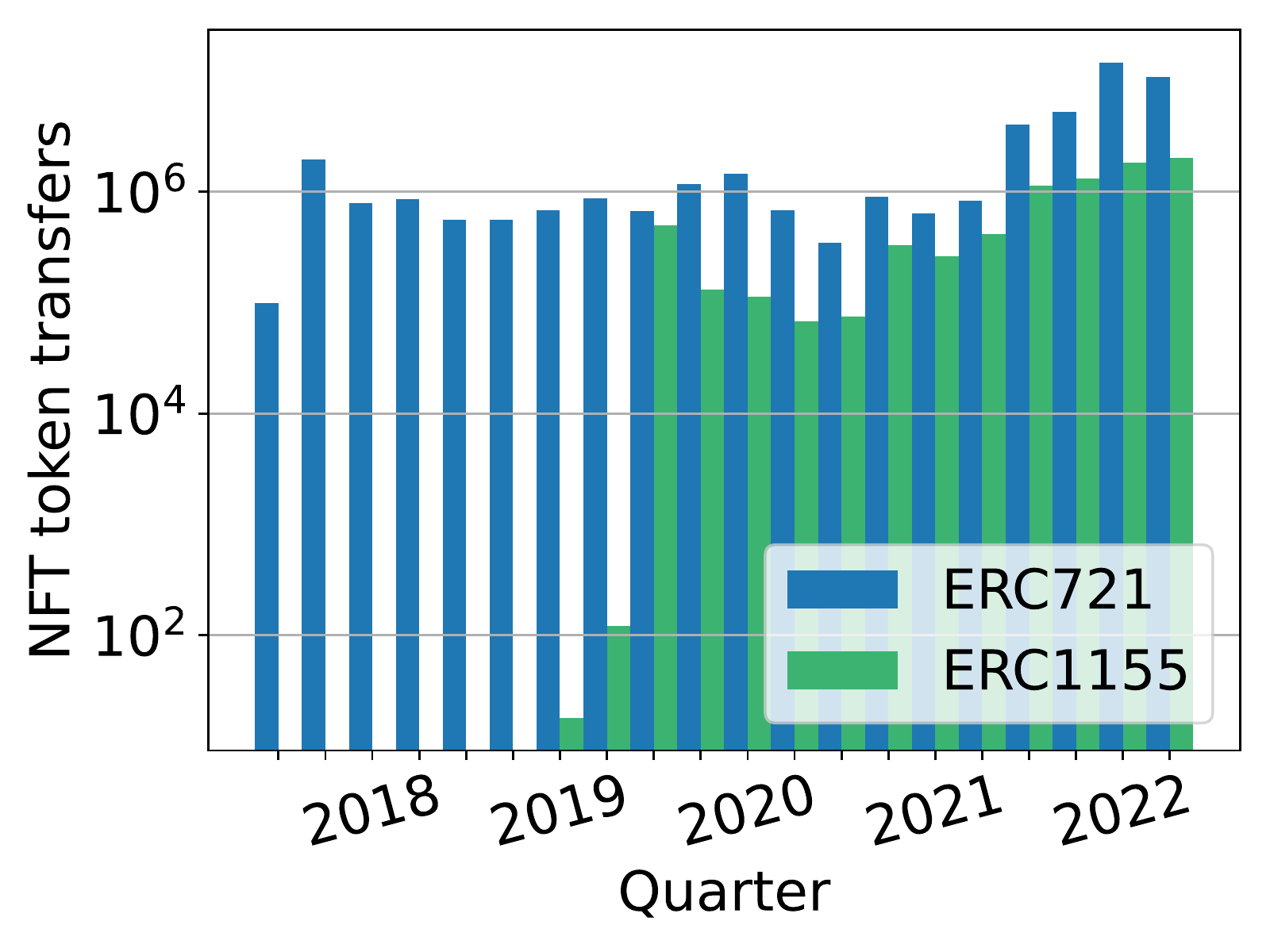}
        \label{fig:ntg_plot_tend_2}
    }
    \caption{Analysis of NTG.}
\end{figure}

For a deeper understanding, we plot the distribution of the indegree (Figure \ref{fig:ntg_plot_indegree_dist}) and outdegree (Figure \ref{fig:ntg_plot_outdegree_dist}) of NTG. 
The indegree of an account in NTG refers to the number of NFTs it received, while the outdegree refers to the number of NFTs it transferred. 
As shown in Figures \ref{fig:ntg_plot_indegree_dist} and \ref{fig:ntg_plot_outdegree_dist}, the distributions of the indegree and outdegree of NTG follow a power law.
By meticulously evaluating the necessary data, we discover that more than half of the transferors (54\%) only receive NFTs once, and 90\% of the transferors buy NFTs for fewer than 14 times. 
In terms of selling, 39\% of transferors sell NFTs once and 90\% sell NFTs for fewer than 28 times, demonstrating that the transferors prefer to sell their NFTs rather than buy them. 
Moreover, we examine the top 10 most important accounts of NTG given by PageRank, and find that they include 6 exchanges, 2 auction contract addresses, and 2 NFT burning addresses.

We also explore the relationships between the trade participants, which demonstrate their trade propensity.
Some metrics of network science are computed, including clustering coefficient, assortativity coefficient, Pearson coefficient, reciprocity coefficient, the number of SCC (strongly connected component) and WCC (weakly connected component) and the size (i.e., how many nodes) of the largest SCC and WCC of NTG. 
The clustering coefficient is very small (i.e., 6E-5), which indicates that there are not conspicuous triangular transfer models in NTG. 
The assortativity coefficient is -0.034, showing that a small-degree node is prone to link with a large-degree node, and vice versa. 
Reciprocity is a measure of the chance that two node in a directed network will be connected to one another.
The reciprocity here is only 0.07, indicating there are a only few bidirectional relations in NFT transfer network. 
Results of these coefficients implies that the trade model in the NFT ecosystem is implicitly suggesting people buy NFTs according to their own biases instead of following suit or click farming. 
The number of SCCs (i.e., 3,588,627) is much greater than the WCC (i.e., 45,845), which shows that a WCC usually contains many SCCs and the NFT transfer among different SCCs should be unidirectional. 
Therefore, an NFT has been transferred from one SCC to another, but it never comes back, which also proves that click farming is rare. 
The size of the largest WCC of NTG is 4,949,725, which stands at 97.5\% of the whole graph, which is similar to some social networks like Twitter, with a value of 94.8\% \cite{ugander2011anatomy}. 

Furthermore, we give an analysis of the trends in the number of NFT traders (Figure \ref{fig:ntg_plot_tend_1}) and NFT transfer times (Figure \ref{fig:ntg_plot_tend_2}). The results exhibit a generally rising trend which is similar to the trend of NCG, indicating that more and more people are participating in the NFT market and trading more and more NFTs.
By comparing Figures \ref{fig:ntg_plot_tend_1} and \ref{fig:ntg_plot_tend_2} with the two trend graphs of NCG (i.e. Figures \ref{fig:ncg_plot_tend_1} and \ref{fig:ncg_plot_tend_2}), we find that the number of creators tends to be larger than the number of transferors in the same quarter, implying that some individuals just try to create their own NFTs without regard for whether they can be sold. 
Thus, the success of the NFT market may be due to the simple attempts of some individual investors who do not know what the NFTs really are. 
These investors may have bought NFTs simply because they saw the popularity and expensive auction prices of NFTs on the Internet.

\subsection{Who Hold These NFTs?}
The holders of NFTs are also an important element of the NFT ecosystem because they represent the final consumer group. Analyzing the owners of NFTs can assist us to determine whether the NFT ecosystem is a bubble or a real boom.

\textbf{NHG Definition and Construction.} 
\emph{NHG = (V, E)}, where \emph{V} is the set of nodes with the outdegree nodes representing accounts and the indegree nodes being NFT IDs, and \emph{E} is the set of directed edges, indicating which accounts hold the NFTs most recently.

\begin{figure}[t]
    \subfigure[NHG]{
        \includegraphics[width=0.47\linewidth]{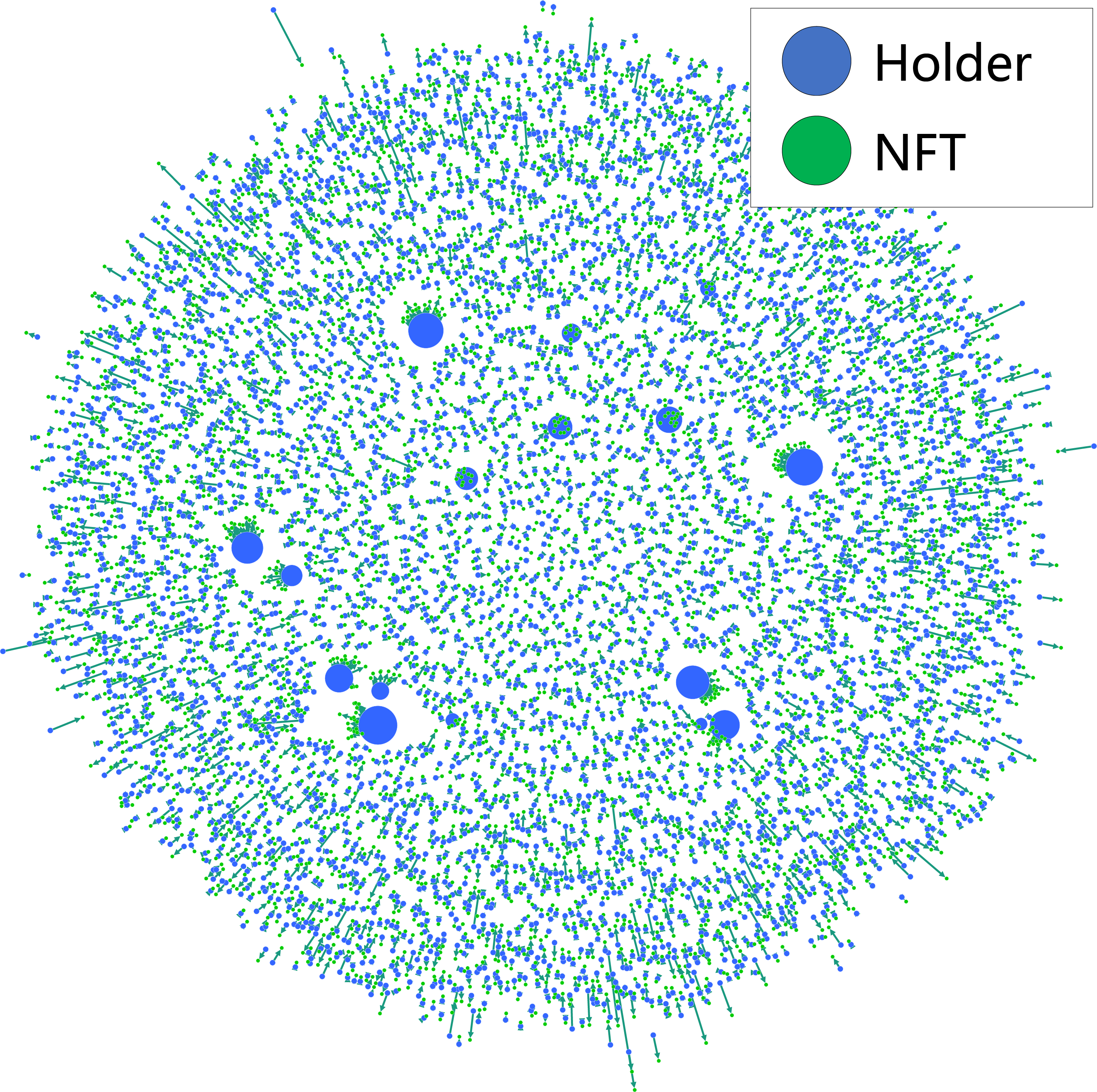}
        \label{fig:nhg}
    }
    \subfigure[Outdegree distribution of NHG]{
        \includegraphics[width=0.47\linewidth]{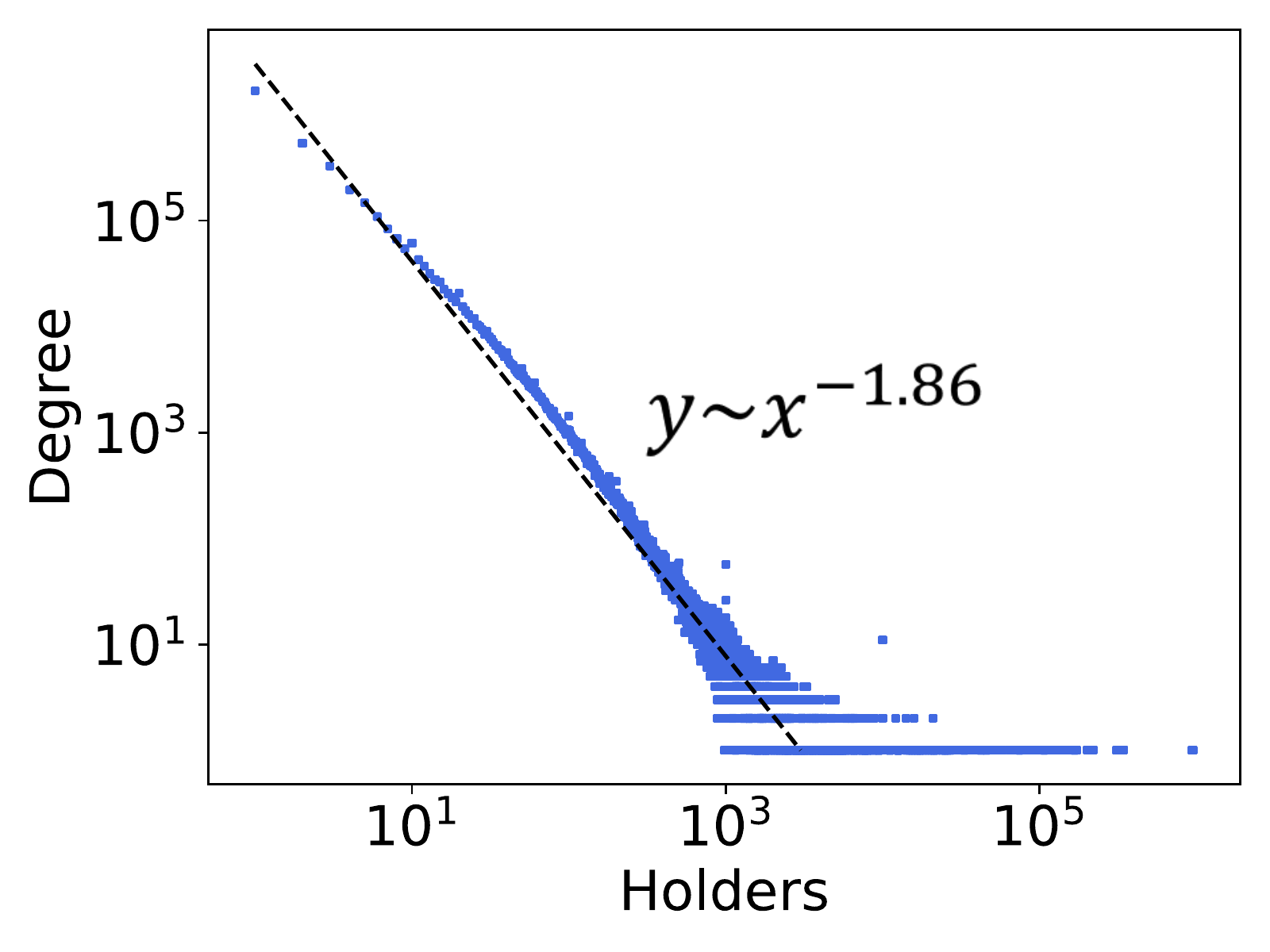}
        \label{fig:nhg_plot_degree_dist}
    }
    \caption{Visualization of NHG and its outdegree distribution.}
\end{figure}

NHG has 78,709,242 nodes and 74,825,956 edges (the same as NCG), which means about 3.88 million accounts (more than NCG) hold near 75 million NFTs. 
The difference between the number of creators and holders implies that NFTs are not only created and sold by creators or insiders but also attract new buyers. 
Furthermore, as with NCG and NTG, the number of holders shows a basic upward trend. 
By examining the edges and nodes belonging to different protocols, we find that 3,616,144 accounts hold ERC721 NFTs and 842,380 accounts hold ERC1155 NFTs and 575,538 accounts hold both kinds of NFTs. 
A sample of NHG with 10,000 edges is showed in Figure \ref{fig:nhg}. 
As can be seen, several accounts hold extensive NFTs which reminds us to compute the distribution of NHG and we show the outdegree distribution of NHG in Figure \ref{fig:nhg_plot_degree_dist}. 

As shown in Figure \ref{fig:nhg_plot_degree_dist}, the outdegree distribution of NHG also follows a power law. 
Approximately 43\% of holders have only one NFT, and 90\% have less than 21 NFTs.
The holders hold about 18 ERC721 NFTs and 0.94 ERC1155 NFT with the total value about 2.75 ETHs on average, which manifest that the head accounts held voluminous NFTs pulling up the average prominently. 
Furthermore, the massive difference in average ownership between ERC721 NFTs and ERC1155 NFTs may be caused by the fact that the number of accounts that collect ERC1155 NFTs is much lower than that of ERC721 NFTs. 
Categorized by the kind of protocols, the ERC721 holding account holds about 20.7 ERC721 (2.9E-7 of total) NFTs and the ERC1155 holding account holds about 4.4 ERC1155 NFTs (1.19E-6 of total) on average, which suggests that ERC1155 NFTs maybe supersaturated. 
The account with the most NFTs is zero address 0x00\footnote{0x0000000000000000000000000000000000000000} who holds 892,766 ERC721 NFTs and 63,134 ERC1155 NFTs pricing at 6779.285 ETHs. 
Among the top 20 accounts with the most NFTs, many accounts do not hold any ERC1155 NFTs. 
Surprisingly, some accounts hold more than 150 thousand NFTs which are only worth less than 1 ETH (less than average), such as 0xd9\footnote{0xd9ab699e5e196139b8a1c8f70ead01b2137fc6a5} indicating the quantity of NFTs that an account owns is not in proportion to the values it has. 

\section{Characteristics of NFTs}
When an NFT is created, transferred, and held, it has the attribute of categories, activeness, and value which are the characteristics that pique the interest of investors and are the decisive factors for investment.
Furthermore, by computing the indicators of these characteristics, we discover some evidences of whether there is any bubble in the NFT ecosystem.
Consequently, we pose a series of questions about these characteristics and attempt to answer them by suggesting new indicators and analyzing the data we collected on chain.
Aiming at finding the differences between NFT categories and determine which category is more worthy of investment, the following researches are carried out in accordance with various categories. 
We use NFT series and single NFT as measurement units to calculate the activeness and value of different NFT categories. 
We discover the following findings while analyzing relevant data.
\begin{enumerate}[-,topsep=0pt]
\item[$\bullet$]\textbf{Finding 1.} In all NFT categories, most NFTs are inactive, and the majority of NFT transfer time are not centralized. There is a large gap in activeness between different NFT categories, and the distribution of leading ENS NFT transfers is anomalous. 
\item[$\bullet$]\textbf{Finding 2.} 
There is a significant value gap between NFTs, even if these NFTs are in the same NFT series. The floor price of an NFT series, especially the ENS, is many orders of magnitude lower than the highest price in the same series. These exorbitant NFT prices may have concealed bubbles in the NFT ecosystem.
\end{enumerate}

\subsection{What is The Classification of NFTs?}
\begin{figure}[t]
    \centering
    \includegraphics[width=0.8\linewidth]{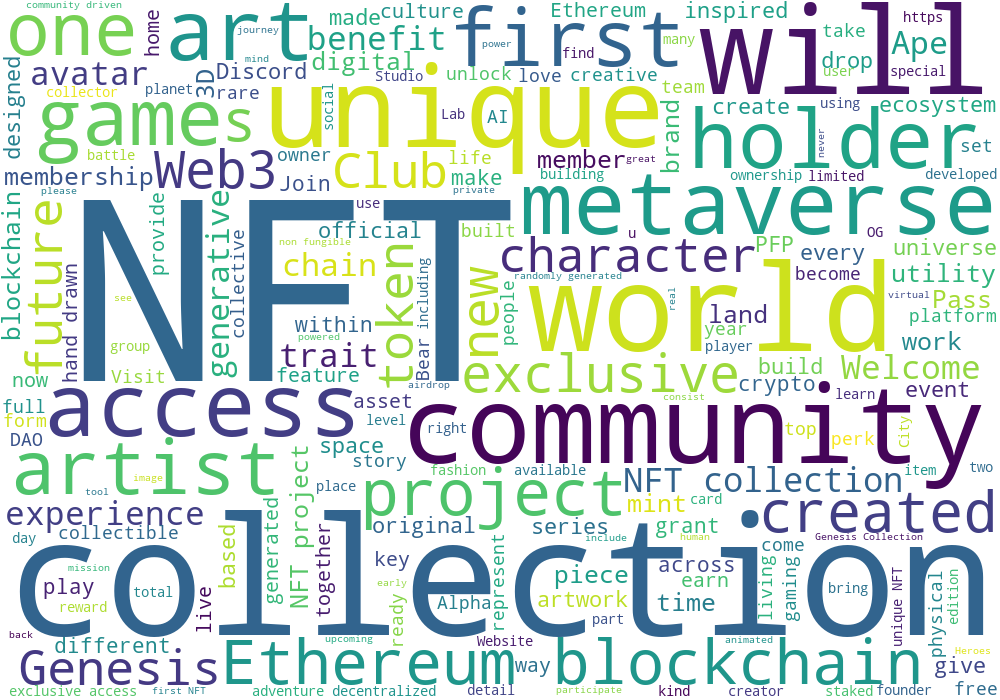}
    \caption{Word cloud for descriptive texts of NFTs.}
    \label{fig:wordcloud}
\end{figure}

According to Etherscan, the categories of NFTs are art, collectibles, ENS (domain names), music, sports, gaming and decentraland, respectively. 
There are about 75 million NFTs in the whole ecosystem, yet only minor NFTs have category labels. 
To get the quantitative features of different NFT categories, we parse the descriptive texts to find the most frequent words used to describe the NFTs and show the word cloud in Figure \ref{fig:wordcloud}. 
As we can see, many categories of NFT appear, such as collection, art, metaverse (i.e., decentraland), and game, which means these kinds of NFT may be created more. 
Note that the classification is based on the NFT series, which means that the NFTs in the same series belong to the same category.

Despite the fact that there are few NFT series with category labels, these NFTs account for the majority of the market value, which are called leading NFTs.
Additionally, the leading NFTs are the bellwethers of the whole NFT market, so the following studies are based on the different categories of these leading NFTs. 
By classifying the leading NFT, we find that among the 15,983,602 NFTs in the 964 NFT series, the gaming has 8,893,870 NFTs in 496 series, and the ENS has 1,602,738 NFTs in only 5 series, with more details shown in the appendix.
We know that in different NFT categories, the number of NFT series and the number of NFTs are not always in direct ratio.
Furthermore, the number of leading NFTs in different categories does not correspond to the total number of NFTs in those categories.

\subsection{How Active are These NFTs?}
\begin{figure*}[t]
    \centering
    \subfigure[The distribution of P value of different NFT categories]{
        \includegraphics[width=0.8\textwidth]{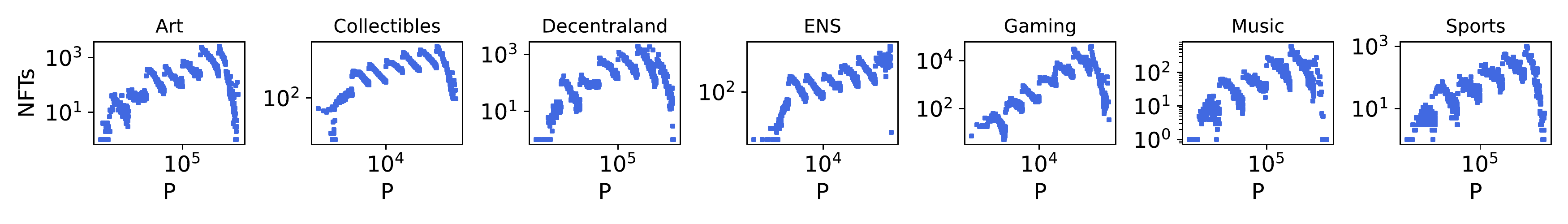}
        \label{fig:activity_P}
    }
    \subfigure[The distribution of fratio of different NFT categories]{
        \includegraphics[width=0.8\textwidth]{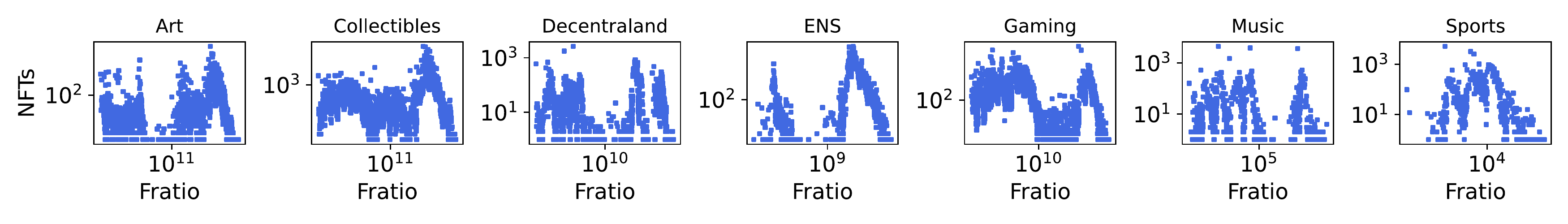}
        \label{fig:activity_fratio}
    }
    \caption{The indicators of NFTs.}
\end{figure*}

Because of the uniqueness of NFTs, it is hard to measure the activeness of an NFT. 
We have noticed that NFTs all begun to become popular with a series. Therefore, we assess the activeness of NFTs in the same series. 
Transformed from the concept of stock trading~\cite{wang1995does}, we propose the turnover ratio to be an indicator measuring the activeness of the NFT series. 
The NFT turnover ratio is calculated by dividing the number of NFT transactions in the series by the total number of NFTs in the series. 
In contrast to the turnover ratio of stocks calculated per year, the NFT turnover ratio is an indicator of the activeness of the NFT series throughout its history. 

Along with the indicator showing the activeness of the NFT series, we also want to know the concentration degree of the transactions of a single NFT which helps us to identify whether a specific NFT is in bubble. 
To this end, we propose an indicator called P value symbolized as $P$ to measure the concentration degree of a specific event like NFT transfers. 
We describe the indicator as 
\begin{equation}
    P=(T_{start}-T_{end}) / N,
    \label{eq:P}
\end{equation}
where $T_{start}$ and $T_{end}$ denote the start time and ending time of the time span we choose to calculate the concentration degree of the transfers of the NFT; \emph{N} is the number of NFT transfers in the chosen time span. 
The start and ending time can be chosen flexibly for different events, but we make them the time of the occurrence of the first and last transfer of this NFT to calculate the $P$ value. 
According to Equation \ref{eq:P}, the higher the P value, the greater the average transfer interval.

As Figure \ref{fig:activity_hfratio_turnover} shows, only a small percentage of NFTs in each NFT category have a high turnover ratio. 
Moreover, in Figure \ref{fig:activity_P}, the distribution of the P value of different NFT categories except ENS follows the pattern more in the middle and less on either side, implying that most NFT transfers are not too concentrated or dispersed. 
By carefully analyzing the results, we discover that ENS has the lowest average turnover and the highest P value suggesting that ENS is the most inactive NFT category.

\subsection{What is the Marketing Performance of NFTs in Different Categories?}
\begin{figure}[t]
	\centering
	\includegraphics[width=0.9\linewidth]{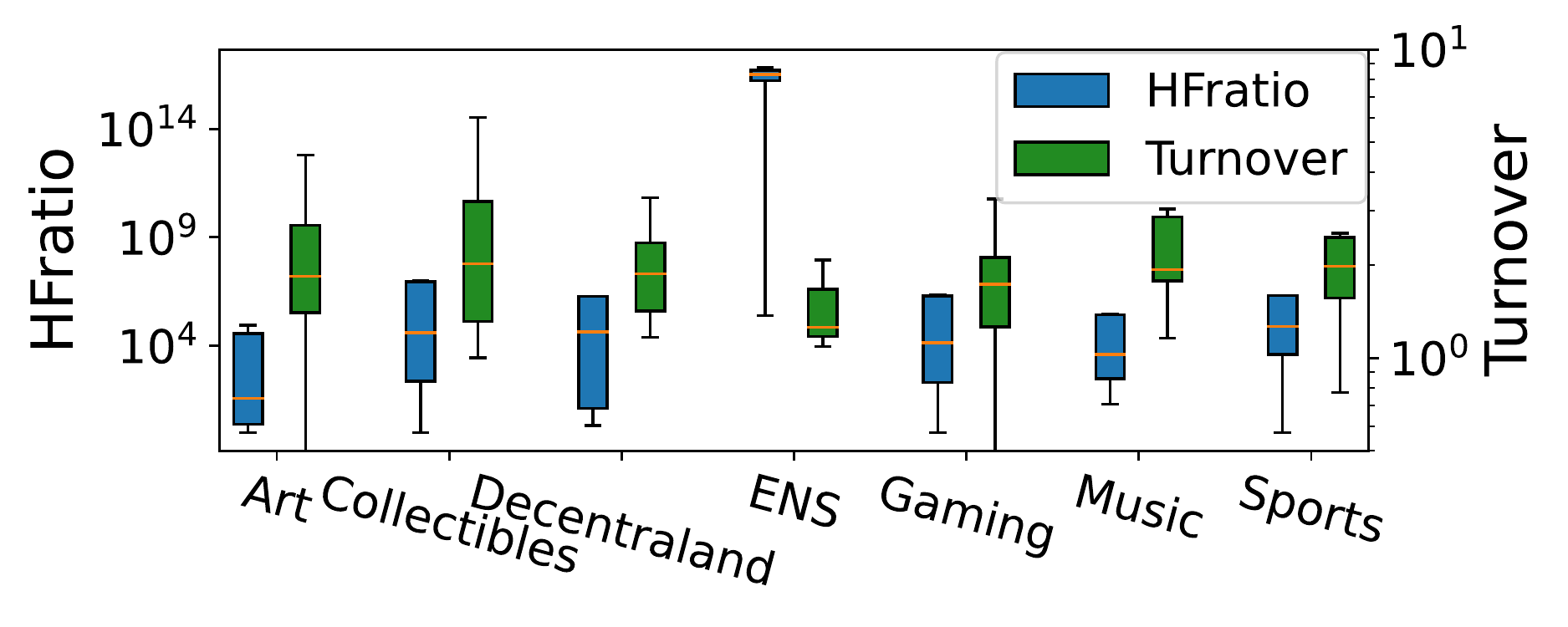}
	\caption{The indicators of NFT series.}
	\label{fig:activity_hfratio_turnover}
\end{figure}

\begin{figure}[t]
	\centering
	\includegraphics[width=0.9\linewidth]{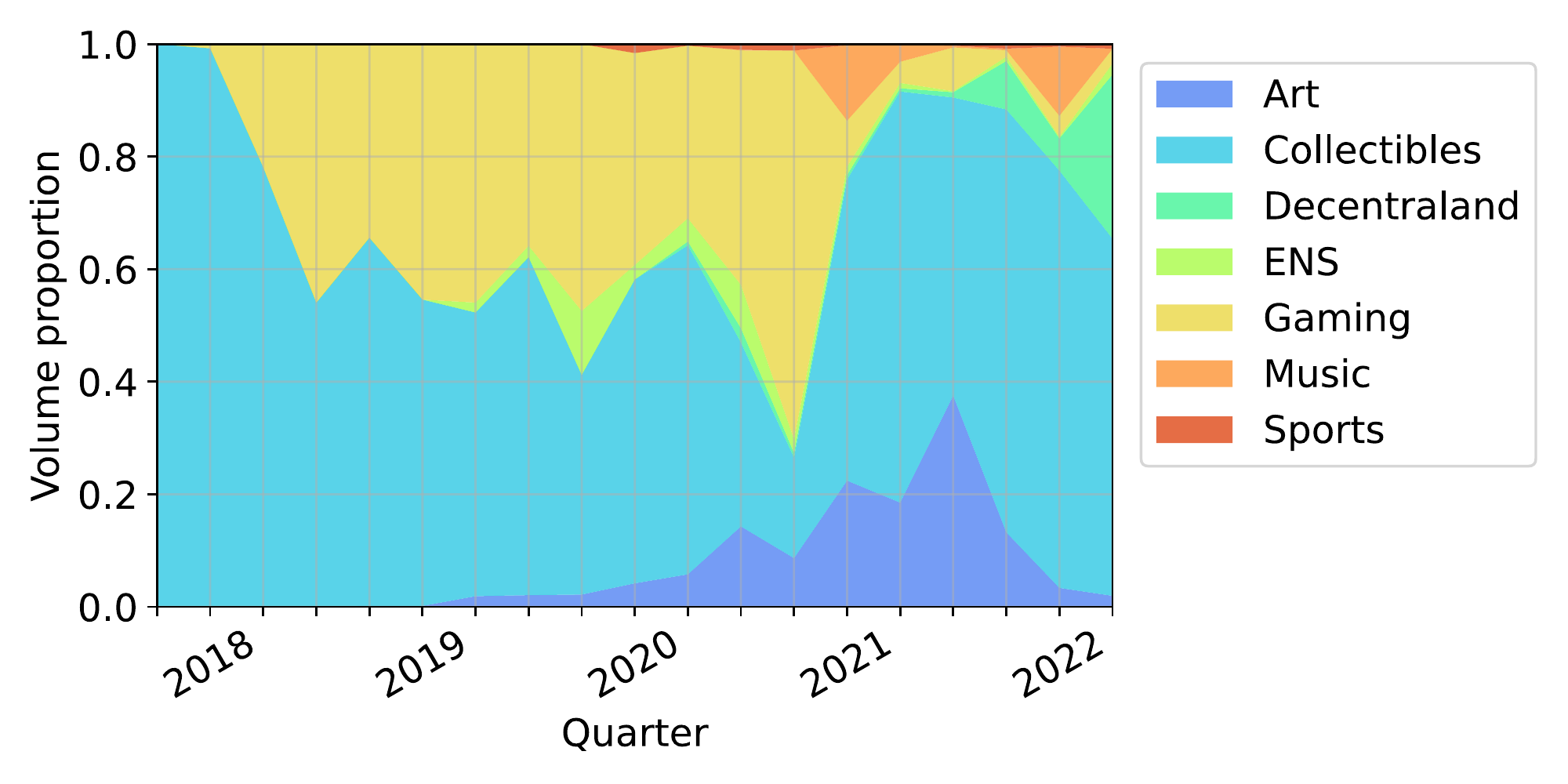}
	\caption{The quarterly volume of different NFT categories.}
	\label{fig:price_volume}
\end{figure}

The marketing performance includes two directions in this paper: market capitalization and volume. 
All other characteristics can be reflected by market capitalization. 
The volume of the NFT reflects the values and activeness in recent times, which is an essential indicator we focus on. 
According the data from the NFT data aggregation platform NFTGO\footnote{https://nftgo.io} combined by our NFT classification, we discover that collectible NFTs account for 73\% of total capitalization, with decentraland, art, and gaming NFTs sharing 10\%, 8\%, and 8\%, respectively. 
The market capitalization proportion of ENS, sports, and music are all below 1\%. 
This result corresponds to our word cloud, which shows that collectible, art, decentraland, gaming NFTs are created more. 
The ranking of market capitalization proportion of leading NFTs grouped by different categories is collectibles (67\%), gaming (11\%), decentraland (9\%), art (8\%), music, ENS and sports, which roughly corresponds to the market capitalization ranking of all NFTs indicating that our choice of NFTs is without loss of generality. 
Though the number of gaming NFTs is larger than collectibles, the market capitalization of collectibles is much higher, suggesting that there may be a large price gap between different NFT categories. 
We show the total market capitalization and average market capitalization of NFT series and single NFTs of different categories in the appendix. 

By analyzing the market capitalization of some specific NFT series, we find that the gap in the same NFT series is also considerably large.
Consequently, we propose an indicator called HFratio which divides the highest price by the floor price in the same NFT series to quantify the price gap in the same NFT series.
By computing the HFratio of the leading NFTs, we find that the HFratio of ENS is much higher than the other categories.
According to the HFratio, turnover ratio, and P value of ENS, we deem that the bubble in ENS NFTs is relatively serious. 
Similarly, to identify the single bubble NFT in the series, we also propose the Fratio, which divides the price of this NFT by the floor price in the same NFT series. 
We plot the distribution of HFratio (Figure \ref{fig:activity_hfratio_turnover}) and Fratio (Figure \ref{fig:activity_fratio}) of different NFT categories.
We discover that while the Fratio distribution of various NFT types is highly varied, it resembles the shape of an ``M'', i.e., there are comparatively more NFTs with high and low Fratio, which may hide bubbles.

To know the recent performance of the leading NFTs, we calculate their quarterly volume from the $4^{th}$ quarter of 2017 to the $2^{th}$ quarter of 2022 shown in Figure \ref{fig:price_volume}. 
The volume of collectibles is often the most except in the $2^{th}$ quarter when the gaming accounts for over 60\% volume. 
However, with the gradual demise of the Covid-19 and the rise of the conception of the metaverse, gaming is replaced by decentraland coming in second place.
This fact shows that volume is not the same as market capitalization but a combination of it and activeness. 
Though the market capitalization of art, game and decentraland is similar, the recent higher activeness makes the decentraland volume the highest among them. 
Therefore, the market capitalization of some NFTs may be just the book value, which cannot be circulating. 
So, we can use volume to find bubbles hidden behind the large market capitalization.

\section{Bubble NFT detection}
\begin{figure}[t]
    \centering
    \includegraphics[width=\linewidth]{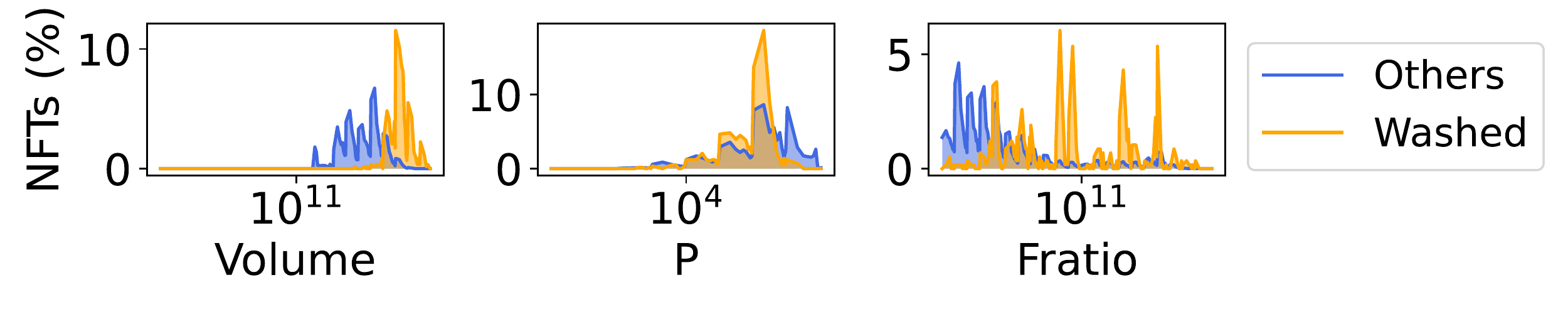}
    \caption{Differences between wash trade and general NFTs}
    \label{fig:detection_wash_feat_dist}
\end{figure}

Prior research revealed that a tiny percentage of NFTs are deviated from generality, with their series having a high turnover ratio or HFratio, and they having a small P value or a high Fratio. 
Experience tells us that bubbles are often hidden from active or valuable NFTs so we think bubble NFTs have some of these anomalies and a high volume. 
The most common way to improve the activeness and value of NFTs is to wash trade. 
However, the users of Ethereum are anonymous, and it is hard to identify the entities behind the accounts. 
Although some addresses are proved belonging to one person, he can create new accounts and new NFTs to seek profits. 
Our approach is to compute the five indicators of NFTs to enclose the bubbles. 
By collecting about 1,309 NFTs with the wash trade label in Dune Analysis\footnote{https://dune.com/cryptok/NFT-Wash-Trading} and calculating these indicators, we find the conspicuous differences between wash trade NFTs and other NFTs, as shown in Figure \ref{fig:detection_wash_feat_dist}. 
We find that the volume and Fratio of wash trade NFTs are generally higher and the P value of wash trade NFTs is lower. 
Moreover, the distributions of these three indicators of wash trade NFTs and others have conspicuous boundaries. 

\begin{algorithm}[t]
    \caption{Detection of Bubble NFTs}
    \label{alg:det_bubble}
    \begin{algorithmic}[1]
        \REQUIRE An NFT series $X$, NTG $g$, and the thresholds $n_1 \sim n_6$.
        \ENSURE A set of bubble NFTs in the series $X$.
        \STATE \textbf{if} X.turnover < $n_1$ or X.HFratio < $n_2$: \textbf{return} $\emptyset$
        \STATE $bubNFTs=\{\}$
        \STATE \textbf{for} x in X.NFTs() \textbf{do}:
        \STATE \ \ \ \ \textbf{if} x.volume < $n_3$ or x.Fratio < $n_4$: \textbf{continue}
        \STATE \ \ \ \ \textbf{if} x.P < $n_5$ or g.count\_transferors(x) < $n_6$:
        \STATE \ \ \ \ \ \ \ \ $bubNFTs = bubNFTs \cup \{x\}$
        \RETURN $bubNFTs$
    \end{algorithmic}
\end{algorithm}

Based on these findings, we propose our new approach to find wash trade NFTs in NFT series, as Algorithm \ref{alg:det_bubble} shows.
The turnover and HFratio are used to describe NFT series, while the volume, Fratio, and P value are used to describe single NFTs. 
In addition, we use a set of thresholds $n_1 \sim n_6$ for the strategy of gradually limiting the scope of bubble detection. 
To begin, we examine the turnover and HFratio of an NFT series to estimate the magnitude of the bubble in this series. 
If the volume and Fratio of this NFT are very high and its P value is modest, suggesting that its activeness and value are extravagant when compared to other NFTs in the same series, we consider it a bubble NFT. 
If the volume and Fratio of the NFT are high but the P value is large, indicating that the transactions of the NFT are not centralized, we examine the number of its transferors to determine whether it is a bubble NFT. 
This approach can save time by looking for exceptions from macro indicators rather than a large quantity of NFT transaction data.

By setting the empirical thresholds $n_1=1.5, n_2=5E+3, n_3=1E+18, n_4=1E+3, n_5=1E+7, n_6=3$, in terms of the above statistics and analysis, we find several bubble NFTs by evaluating the leading NFTs using our approach and choosing two epitomes of NFT, which are reported in the wash trade, to demonstrate the feasibility of our approach. 
The level 7 at \{12, 20\}\footnote{\url{looksrare.org/collections/0x4e1f41613c9084fdb9e34e11fae9412427480e56/1865}}, a NFT belonging to the NFT series called ``Terraforms by Mathcastles" with a turnover ratio of 3.53 and an HFratio of 2.0E+13, has a volume of over 3E+21 wei, a P value of 8.1E+3, and a Fratio of 3.0E+12. 
Because the indications of activeness and value of this NFT are all quite higher than general NFTs, we conclude that it is a bubble NFT.
Additionally, we discover a bubble NFT, the emoji ENS\footnote{\url{looksrare.org/collections/0x57f1887a8bf19b14fc0df6fd9b2acc9af147ea85/6240067598452542859452315860854899245467171159805616565629252684328784735464}}, with a turnover of 1.26 and an HFratio of 6.6E+16 in its series and having a volume of 1E+19 wei, a Fratio of 12499, and a P value of 2.05E+7. 
Since the P value is high, it is simple to locate the transferors. 
As a result, we discover that there are just 2 transferors and 4 transfers of this NFT in NTG. 
Due to its high activeness, value, and anomalous bidirectional transfers, we classify this NFT as a bubble NFT.

\section{conclusion and future work}
We conducted systematic research to characterize the ERC721 and ERC1155 NFT ecosystems. 
By using the Ethereum data services, we collected all NFT token transfer records on Ethereum and then constructed three graphs to recognize the characteristics of NFT traders. 
Combining the data and labels, we calculated the category, activeness, and value characteristics of NFTs. 
Moreover, we proposed new indicators to measure the activeness of NFTs and a new approach to find bubble NFTs, especially wash trade NFTs. 
After these analyses, we made the following summaries of the NFT market:
i) Most of the users are inactive and hold a few NFTs. But the users that do frequent transactions and have a large number of NFTs are more likely to engage in the trade of bubble NFTs.
ii) Although most NFTs are inactive and worthless, it is the common state in the market economy. The NFTs with high activeness and market capitalization may be in bubbles.
iii) The prosperity of the NFT ecosystem will persist for some time due to the rise in the number of creators, transferors, and holders. Nevertheless, some bubbles will be enclosed for the reason that some whales or scammers rig the price by wash trade or hype.
In the future, we plan to enhance our bubble NFT detection algorithm by adding more comprehensive characteristics and making it capable of detecting additional types of bubble NFTs like scam NFTs and NFTs created by automatic programs. 
By using the bubble NFT labels, we can analyze the characteristics and trade patterns of the bubble NFT traders to give better guidance for NFT investment. 


\bibliographystyle{ACM-Reference-Format}
\bibliography{sample-base}

\newpage
\section{appendix}
\begin{table}[ht]
  \begin{center}
  \caption{The numbers of leading NFT series and leading NFTs in different NFT categories}
    \begin{tabular}{l|l|l} 
      \toprule
      \textbf{NFT Category} &  \textbf{Leading series} &
      \textbf{Leading NFT}\\
      \hline
      Decentraland &35&251,902\\
      Gaming&496& 8,893,870\\
      Music&58& 51,625\\
      ENS&5& 1,602,738\\
      Art&129&256,056\\
      Collectibles&189&4,840,699\\
      Sports&52&86,742\\
      \bottomrule
    \end{tabular}
  \end{center}    
\end{table}
\begin{table}[ht]
  \begin{center}
  \caption{The market capitalization of different NFT categories as a whole, as the average by leading NFT series, and as the average by leading NFTs.}
    \begin{tabular}{l|l|l|l} 
      \toprule
      \makecell[l]{NFT\\category}&
      \makecell[l]{Market cap\\by Ether}&
      \makecell[l]{Average\\market cap of\\leading series}&
      \makecell[l]{Average\\market cap of\\leading NFTs}\\
      \hline
      Decentraland&915,154&26,147&3.63\\
      Gaming&1,112,254&2,242&0.12\\
      Music&265,909&4,584&5.15\\
      ENS&163,669&32,733&0.10\\
      Art&768,564&5,957&3.00\\
      Collectibles&6,614,524&34,997&1.36\\
      Sports&43,726&840&0.50\\
      \bottomrule
    \end{tabular}
  \end{center}    
\end{table}

\appendix

\end{document}